\documentstyle[preprint,tighten,aps,epsfig,colordvi,psfig]{revtex}

\def\lsim{\mathrel{\rlap{\lower4pt\hbox{\hskip1pt$\sim$}}
    \raise1pt\hbox{$<$}}}                
\def\gsim{\mathrel{\rlap{\lower4pt\hbox{\hskip1pt$\sim$}}
    \raise1pt\hbox{$>$}}}                

\begin{document}
\setlength{\textheight}{23cm}
\draft
\title{Gauge theories with overlap fermions \\ in an arbitrary representation: \\ Evaluation of the 3-loop $\beta$-function}
\author{M. Constantinou, H. Panagopoulos\\}
\address{
Department of Physics, University of Cyprus,\\
P.O.Box 20537, Nicosia CY-1678, Cyprus\\
{\it email:}{ phpgmc1@ucy.ac.cy, haris@ucy.ac.cy}}

\maketitle

\begin{abstract}
This work presents the calculation of the relation between the bare
coupling constant $g_0$ and the $\rm \overline{MS}$-renormalized
coupling $g_{\,\rm \overline{MS}}$, $g_0 = Z_g(g_0,a\mu) g_{\,\rm
  \overline{MS}}$, to 2 loops in perturbation theory, with fermions in an arbitrary
representation of the gauge group $SU(N)$. Our calculation is
performed using overlap fermions and Wilson gluons, and the background
field technique has been chosen for convenience. The corresponding results
in the fundamental representation appear in our longer publication~\cite{CP}. 

The 3-loop coefficient of the bare $\beta$-function, $b_2^L$, is
extracted using the 2-loop expression for $Z_g$, and it is presented
as a function of the overlap parameter
$\rho$, the number of fermion flavors ($N_f$) and the number of colors ($N$). We also provide the
expression for the ratio $\Lambda_L/\Lambda_{\,\overline{\rm MS}}$, in
an arbitrary representation. A plot of $\Lambda_L/\Lambda_{\,\overline{\rm MS}}$ 
is given in the adjoint representation.



\end{abstract}
\newpage

\section{INTRODUCTION}

In this letter we present a computation of the bare $\beta$-function in
the lattice formulation of the $SU(N)$ gauge theory, with $N_f$ flavors of overlap fermions
in an arbitrary representation of the gauge group $SU(N)$.
This work is a sequel to a longer publication~\cite{CP}, in which the
computation was performed with fermions in the fundamental
representation. We will refer to~\cite{CP} for the general background
and notation.

While the main application of $SU(N)$ gauge theories on the lattice
regards QCD, where fermions are in the fundamental representation of
$SU(3)$, there has recently been some interest in gauge theories with
fermions in other representations and with $N\neq 3$. Such theories
are being studied in various contexts~\cite{KUY,GG,CDDLP,C,BPV,S,GK,EHS}, e.g., supersymmetry,
phase transitions, and the 'AdS/QCD' correspondence.
 
In later years, use of non-ultralocal actions which preserve chiral
symmetry on the lattice has become more viable for numerical simulations.
The two actions which are being used most frequently are overlap
fermions~\cite{NN,Neuberger-98,Neuberger-01} 
based on the Wilson fermion action and domain-wall fermions~\cite{Kaplan,FS}.

Overlap fermions are notoriously difficult to study, both numerically
and analytically. 
Many recent promising investigations involving simulations with
overlap fermions have appeared; see, e.g.,
Refs.~\cite{DLS,DMZACDHLLT,BJSSS,BGHHLR,JLQCD,BFLW,IKKSSW}.  
Regarding analytical computations, the only ones performed thus far
have been either up to 1 loop, such as
Refs.~\cite{APV,AFPV,CG,Capitani,D,GHPRSS,IP},
or vacuum diagrams at higher loops~\cite{AP,SP}.
{\it{The present work (along with Ref.~\cite{CP}) is the first one involving non-vacuum diagrams
    beyond the 1-loop level.}}

We compute the 2-loop renormalization $Z_g$ of the bare lattice
coupling constant $g_0$ in the presence of overlap fermions.
We relate $g_0$ to the renormalized coupling constant $g_{\,\rm
  \overline{MS}}$ as defined in the ${\rm \overline{MS}}$ scheme
at a scale $\bar{\mu}$; at large momenta, these quantities are related as follows
\begin{equation}
\alpha_{\rm \,\overline{MS}}(\bar{\mu}) = \alpha_0 + d_1(\bar{\mu} a)\alpha_0^2 + d_2(\bar{\mu} a)\alpha_0^3 + ... \,,
\end{equation}
($\alpha_0=g_0^2/4\pi, \, \alpha_{\rm \,\overline{MS}}=g_{\rm \,\overline{MS}}^2/4\pi, \, a:$ lattice spacing).
The 1-loop coefficient $d_1(\bar{\mu} a)$ has been known for a long time; several evaluations of $d_2(\bar{\mu} a)$ 
have also appeared in the past $\sim$10 years, either in the absence of fermions \cite{LW,AFP}, or using the Wilson 
\cite{CFPV} or clover \cite{BP,BPP} fermionic actions. Knowledge of
$d_2(\bar{\mu} a)$, together with the 3-loop 
$\rm \overline{MS}$-renormalized $\beta$-function \cite{TVZ}, allows
us to derive the 3-loop bare lattice $\beta$-function, 
which dictates the dependence of the lattice spacing on $g_0$. 
In particular, it provides a correction to the standard 2-loop
asymptotic scaling formula which defines the renormalization group invariant scale $\Lambda_L$.
Ongoing efforts to estimate the running coupling from the lattice \cite{Sommer,ALPHA,KKPZ,MTDFGLNS} 
have relied on a mixture of perturbative and non-perturbative investigations. As a particular example, relating 
$\alpha_{\rm \,\overline{MS}}$ to $\alpha_{\rm SF}$ (SF: Schr\"odinger
Functional scheme), entails an intermediate passage through the bare coupling $\alpha_0$; the conversion
from $\alpha_{\rm \,\overline{MS}}$ to $\alpha_0$ is then carried out perturbatively.

The present study, being the first of its kind in calculating 2-loop diagrams with overlap vertices and external momentum dependence,
had a number of obstacles to overcome. One first complication is the size of the algebraic form of the Feynman vertices;
as an example, the vertex with 4 gluons and a fermion-antifermion pair contains $\sim$724,000 terms when expanded.
Upon contraction these vertices lead to huge expressions (many millions of terms); this places severe requirements 
both on the necessary computer RAM and on the efficiency of the
computer algorithms which we must design to manipulate such
expressions automatically.
Numerical integration of Feynman diagrams over loop momentum variables is performed on a range of lattices, with finite size $L$, 
and subsequent extrapolation to $L\rightarrow \infty$. As it turns out, larger $L$ are required for an accurate
extrapolation in the present case, as compared to ultra-local actions.
In addition, since the results depend nontrivially on the parameter $\rho$ of the overlap action, numerical evaluation must be
performed for a sufficiently wide set of values of $\rho$, with an almost proportionate increase in CPU time.
Extreme choices of values for $\rho$ ($\rho\gsim 0,\, \rho\lsim 2$) show unstable numerical behaviour, which is attributable to the spurious poles of the fermion propagator
for these choices; this forces us to even larger $L$. A consequence of all complications noted above is an extended use of CPU time:
Our numerical integration codes, which ran on a 32-node cluster of dual CPU Pentium IV processors, required a total of $\sim$50 years of CPU time.

The paper is organized as follows:
The necessary theoretical background for the $\beta$-function, $Z_g$ and
$\Lambda_L/\Lambda_{\,\overline{\rm MS}}$ is given in Section II.
Section III describes briefly the overlap action and Section IV 
provides details on our computation and results for the fundamental
representation. A generalization of the above results to an arbitrary
representation is carried out in Section V. 
We employ the formulae of Section V to find 
numerical results for the 3-loop $\beta$-function and the ratio
$\Lambda_L/\Lambda_{\,\overline{\rm MS}}$. 
Results for the adjoint representation appear in Section VI.  


\section{THEORETICAL BACKGROUND}

For the lattice regularization a bare $\beta$-function is defined as
\begin{equation}
\displaystyle \qquad\qquad  \beta_L(g_0)= -a{dg_0\over da} \Bigg|_{g,\,\bar{\mu}}
\end{equation}
where $\bar{\mu}$ is the renormalization scale and $g$ $(g_0)$ the
renormalized (bare) coupling constant.
It is well known that in the asymptotic limit for QCD ($g_0\rightarrow 0$), one can write the expansion of the $\beta$-function
in powers of $g_0$, that is
\begin{equation}
\beta_L(g_0) =-b_0 \,g^3_0 -b_1 \,g_0^5 - b_2^{L}\,g_0^7 - ... 
\end{equation}
The coefficients $b_0, b_1$ are universal, regularization independent
constants (see Eqs.~(\ref{univ_coeff_b0}), (\ref{univ_coeff_b1})).
On the contrary, $b_i^{L}$ ($i \ge 2 $) depends on the regulator; it must be determined perturbatively.
In the present work we calculate the coefficient $b_2^{L}$ using the $\rm overlap$ fermionic action and Wilson gluons.

$\displaystyle \beta_L(g_0)$ can be computed from the renormalization function 
$Z_g$, which is defined through $g_0 = Z_g^{L,\overline{\rm MS}}(g_0,a\bar{\mu})
g\,$, and relates the bare lattice coupling to the one renormalized in
the $\overline{\rm MS}$ scheme. Up to 2-loops, $Z_g$ can be written in the form
\begin{eqnarray}
&&(Z_g^{L,\overline{\rm MS}}(g_0,a\bar{\mu}))^2 = 1 + g_0^2\,(2b_0 \ln(a\bar{\mu})+ l_0) +g_0^4\,(2b_1 \ln(a\bar{\mu})+ l_1) + O(g_0^6) 
\label{Zg}
\end{eqnarray}
The quantities $b_0, b_1$ and $l_0$ have been known in the literature 
for quite some time~\cite{TVZ,APV}; $b_0$ and $b_1$ are the same as
those of the renormalized $\beta$-function. 

The constant $l_0$ is related to the ratio of the $\Lambda$ parameters associated with 
the particular lattice regularization and the $\overline{\rm MS}$ renormalization scheme
\begin{equation}
\displaystyle \Lambda_L/\Lambda_{\,\overline{\rm MS}} \equiv e^{l_0/(2b_0)}
\label{def_L}
\end{equation}
For overlap fermions in the fundamental representation the exact form of $l_0$ has been given in Ref.~\cite{APV}.
The value of $l_0$ for different representations is presented in Section V.

One can extract $b_2^L$ from $Z_g$ through
\begin{equation}
b_2^L= b_2 -b_1l_{0}+ b_0 l_{1}
\label{b2lrel}
\end{equation}
where $b_2$ is the 3-loop coefficient of the renormalized
$\beta$-function in the $\overline{\rm MS}$ scheme~\cite{TVZ}.
Thus, the evaluation of $b_2^{L}$ requires only the determination of the 2-loop quantity $l_{1}$.

The most convenient and economical way to proceed with calculating $Z_g(g_0,a\bar{\mu})$ 
is to use the background field technique \cite{Abbott,EM,LW2}, 
in which the following relation is valid
\begin{equation}
Z_A(g_0,a\bar{\mu})   Z_g^2(g_0, a\bar{\mu}) = 1
\end{equation}
where $Z_A$, defined as: $A^{\mu}(x) = Z_A(g_0,a\bar{\mu})^{1/2} A_{R}^{\mu}(x)$ 
is the background field renormalization function ($A^{\mu}$ ($A_R^{\mu}$): bare (renormalized) background field). 
In this framework, instead of $Z_g(g_0,a\bar{\mu})$, it suffices to compute $Z_A(g_0,a\bar{\mu})$,
with no need to evaluate any 3-point functions.
For this purpose, we consider the background field one-particle irreducible (1PI) 2-point function 
at momentum $p$, both in the continuum and on the lattice. In the notation of
Ref.~\cite{LW}, the lattice 2-point function
$\Gamma^{AA}_L(p)^{ab}_{\mu\nu}$ can be expressed in terms of the scalar function $\nu(p)$
\begin{equation}
\sum_\mu \Gamma^{AA}_L(p)^{ab}_{\mu\mu} =  
-\delta^{ab}3\widehat{p}^2 
\left[ 1 - \nu(p)\right]/g_0^2, \quad\quad\quad
\nu(p) = g_0^{2} \nu^{(1)}(p)+g_0^{4} \nu^{(2)}(p)+... 
\label{2pt-functionA}
\end{equation}
($\hat{p}_\mu=(2/a) \sin(ap_\mu/2)$). 
There follows
\begin{equation}
Z_A = {1 - \nu_R(p,\bar{\mu},g)\over 1 - \nu(p,a,g_0)}
\end{equation}
where $\nu_R$ (as well as $\omega_R$ below) is the continuum counterpart of $\nu$
($\omega$) (in dimensional regularization, $\rm \overline{MS}$ subtraction).
The gauge parameter $\lambda$ must also be renormalized (up to 1 loop), in order to compare lattice and continuum results
\begin{eqnarray}
\lambda = Z_Q\lambda_0 \, , \qquad Z_Q = 1+ g_0^2 z_Q^{(1)}+...
\end{eqnarray}
($Z_Q$: renormalization function of the quantum field). The coefficient $z_Q^{(1)}$ is obtained from the 
quantum field 1PI 2-point function on the lattice ($\Gamma^{QQ}_L(p)^{ab}_{\mu\nu}$) through
\begin{equation}
\sum_\mu \Gamma^{QQ}_L(p)^{ab}_{\mu\mu} =
-\delta^{ab}\widehat{p}^2
\left[ 3\left( 1 - \omega(p)\right) + \lambda_0\right], \quad\quad\quad
\omega(p) = g_0^{2}\, \omega^{(1)}(p)+{\cal O}(g_0^{4})
\label{2pt-functionQ}
\end{equation}
\vspace{-.45cm}
\begin{equation}
z_Q^{(1)} = \omega^{(1)}(p,a,g_0)  - \omega_R^{(1)}(p,\bar{\mu},g)
\end{equation}
In terms of the perturbative expansions Eqs.~(\ref{2pt-functionA}), (\ref{2pt-functionQ}),
$Z_g^2$ becomes
\begin{equation}
Z_g^2 = \Big[ 1 + g_0^2\, (\nu_R^{(1)} - \nu^{(1)}) + g_0^4\, (\nu_R^{(2)} - \nu^{(2)}) + 
\lambda_0\, g_0^4\, (\omega^{(1)} - \omega_R^{(1)})\frac{\partial \nu_R^{(1)}}{\partial \lambda} \Big]_{\lambda=\lambda_0}
\label{Zg_expanded}
\end{equation}
Since the quantities of interest are gauge invariant, we choose to work in the bare Feynman gauge, $\lambda_0=1$, for convenience.
In order to compute $Z_A$ we need the expressions for $\nu_R^{(1)},\,\nu_R^{(2)},\,z_Q^{(1)},\,\nu^{(1)},\,\nu^{(2)}$. 
The $\overline{\rm MS}$ renormalized functions  $\nu_R,\,\omega_R$,
necessary for this calculation to 2 loops can be found in Refs.~\cite{LW} ($N_f=0$) and \cite{CFPV} (arbitrary $N_f$).

For the lattice quantities, the gluonic contributions ($N_f = 0$) have been presented in previous publications~\cite{LW,AFP} 
(for the Wilson gauge action) 
\begin{eqnarray}
\omega^{(1)}(p,\lambda_0=1) =&& -{5N\over 48\pi^2}\ln{(a^2p^2)} -{1\over 8N}+0.137286278291N 
\label{omega1_1} \\
\nonumber \\
\nu^{(1)}(p,\lambda_0=1) =&& -{11N\over 48\pi^2}\ln{(a^2p^2)} -{1\over 8N} + 0.217098494367N 
\label{nu1_1} \\ 
\nonumber \\
\nu^{(2)}(p,\lambda_0=1) =&& -{N\over 32\pi^4}\ln{(a^2p^2)} +{3\over 128N^2}-0.01654461954+0.0074438722N^2 
\label{nu2_1} 
\end{eqnarray}

The fermionic diagrams that contribute to $\nu^{(1)}$ (1-loop) and
$\nu^{(2)}$ (2-loop) appear in Fig.1 and Fig.2 of Ref.~\cite{CP}, where
$\nu^{(2)}$ was perturbatively calculated for the first time using overlap fermions and Wilson gluons.
The fermion part of $\omega^{(1)}$ coincides with that of $\nu^{(1)}$. 
For overlap fermions, there are no diagrams with a mass counterterm, by
virtue of the exact chiral symmetry of the overlap action.


\section{OVERLAP ACTION}

In recent years, overlap fermions are being used ever more extensively in numerical simulations,
both in the quenched approximation and beyond. 
This fact, along with the desirable properties of the overlap action, was our motivation 
to calculate the $\beta$-function with this type of fermions. 
The important advantage of the overlap action is that it preserves chiral symmetry 
while avoiding fermion doubling. It is also ${\cal O}(a)$ improved. 
The main drawback of this action is that it is necessarily non-ultralocal; as a consequence,
both numerical simulations and perturbative studies are extremely difficult and demanding 
(in terms of human, as well as computer time).

The overlap action is given by~\cite{Neuberger-98}
\begin{equation}
S_{\rm overlap} = a^8 \sum_{n,m} \bar{\Psi}(n) \, D_N (n,m) \, \Psi(m)
\end{equation}
where $D_N (n,m)$ is the overlap-Dirac operator
\begin{eqnarray}
D_N (n,m) &=& \rho \Bigg[\frac{\delta_{n,m}}{a^4}-\left(X\frac{1}{\sqrt{X^\dagger X}}\right)_{nm}\Bigg], \qquad
X = \frac{1}{a^4} \left(D_W -\rho \right)
\end{eqnarray}
and $D_W$ is the Wilson-Dirac operator (with the Wilson parameter $r$
set to 1)
\begin{equation}
D_{\rm W} = {1\over 2} \left[ \gamma_\mu \left( \nabla_\mu^*+\nabla_\mu\right)
 - a\nabla_\mu^*\nabla_\mu \right], \qquad
\nabla_\mu\psi(x) = {1\over a} \left[ U(x,\mu) \psi(x + a\hat{\mu})- \psi(x)\right]
\end{equation}
The overlap parameter $\rho$
is restricted by the condition $0<\rho <2$ to guarantee the correct
pole structure of $D_N$. The coupling constant is included in the link
variables, present in the definition of $X$, and one must study the
perturbative expansion of $D_N$ in powers of $g_0$. The expansion of $X$ in
momentum space reads 
\begin{equation}
X(p',p)=\underbrace{\chi_0(p)(2\pi)^4 \delta_P(p' - p)}_{tree-level} + 
\underbrace{X_1(p',p)+X_2(p',p)}_{1-loop} +
\underbrace{X_3(p',p)+X_4(p',p)}_{2-loop} + O(g^5_0)
\end{equation}
where $\chi_0$ is the inverse fermion propagator and $X_i$ are the vertices of the Wilson fermion action 
with $i$ gluons ($p$ ($p'$): fermion (antifermion) momentum). 
The construction of all overlap vertices relevant to the present 2-loop computation 
make use of $\chi_0$ and $X_1-X_4$; these quantities can be written in the compact form
\begin{eqnarray}
&&\chi_0(p)=\frac{i}{a}\sum_{\mu}\gamma_{\mu}\sin(ap_{\mu})+\frac{1}{a}\sum_{\mu}\Big(1-\cos(ap_{\mu})\Big)-\frac{\rho}{a} \nonumber \\
&&X_1(p',p)=g_0\int
d^4k\delta(p'-p-k)\sum_{\mu}A_{\mu}(k)V_{1,\mu}\Big(\frac{p'+p}{2}\Big) \nonumber \\
&&X_2(p',p)=\frac{g_0^2}{2}\int
\frac{d^4k_1d^4k_2}{(2\pi)^4}\delta(p'-p-k_1-k_2)\sum_{\mu}A_{\mu}(k_1)
A_{\mu}(k_2)V_{2,\mu}\Big(\frac{p'+p}{2}\Big) \nonumber\\
&&X_3(p',p)=\frac{g_0^3}{3!}\int
\frac{d^4k_1d^4k_2d^4k_3}{(2\pi)^8}\delta(p'-p-\sum_{i=1}^3 k_i)\sum_{\mu}\prod_{i=1}^3 A_{\mu}(k_i)
\Big[-a^2 V_{1,\mu}\Big(\frac{p'+p}{2}\Big) \Big] \nonumber\\
&&X_4(p',p)=\frac{g_0^4}{4!}\int
\frac{d^4k_1d^4k_2d^4k_3d^4k_4}{(2\pi)^{12}}\delta(p'-p-\sum_{i=1}^4 k_i)\sum_{\mu}\prod_{i=1}^4 A_{\mu}(k_i)
\Big[-a^2 V_{2,\mu}\Big(\frac{p'+p}{2}\Big) \Big]
\label{X's}
\end{eqnarray}
where
\begin{eqnarray}
V_{1,\,\mu}(p)=i\,\gamma_{\,\mu}\cos(ap_{\,\mu})+\sin(ap_{\,\mu}), \quad V_{2,\,\mu}(p)=-i\,\gamma_{\,\mu}a\,\sin(ap_{\,\mu})+a\,\cos(ap_{\,\mu})
\end{eqnarray}
In Eqs.~(\ref{X's}) $A_{\,\mu}$ represents a gluon field. The use of the background field 
technique implies that instead of the generic gluon fields, 
one must consider all possible combinations of background 
(\Red{A}) and quantum (\Blue{Q}) fields, which are generated through
the replacement
\begin{equation}
e^{i a g_0 A_{\mu}(x)} \rightarrow e^{i a g_0 Q_{\mu}(x)}\cdot e^{i a A_{\mu}(x)}
\end{equation}
For example,
\begin{eqnarray}
X_3(p',p)= X_3^{\Blue{Q}\Blue{Q}\Blue{Q}}(p',p)+X_3^{\Blue{Q}\Blue{Q}\Red{A}}(p',p)+
X_3^{\Blue{Q}\Red{A}\Red{A}}(p',p)+X_3^{\Red{A}\Red{A}\Red{A}}(p',p)
\label{X3}
\end{eqnarray}
where $X_3^{\Blue{Q}\Blue{Q}\Red{A}}$ is given by
\begin{eqnarray}
X_3^{\Blue{Q}\Blue{Q}\Red{A}}(p',&&p)= \frac{g_0^2}{4}\int
\frac{d^4k_1d^4k_2d^4k_3}{(2\pi)^8}\delta(p'-p-k_1-k_2-k_3)\times \nonumber \\
&&\sum_{\mu}\Bigg[-a^2 V_{1,\mu}\Big(\frac{p'+p}{2}\Big)\Big(Q_{\mu}(k_1)Q_{\mu}(k_2)A_{\mu}(k_3)+A_{\mu}(k_3)Q_{\mu}(k_2)Q_{\mu}(k_1)\Big) \nonumber \\
&&\phantom{spa\Bigg[}+i a V_{2,\mu}\Big(\frac{p'+p}{2}\Big)\Big(Q_{\mu}(k_1)Q_{\mu}(k_2)A_{\mu}(k_3)-A_{\mu}(k_3)Q_{\mu}(k_2)Q_{\mu}(k_1)\Big) \Bigg]
\end{eqnarray}
and similarly for the remaining terms of Eq.~(\ref{X3}).

At this point we can proceed with the perturbative expansion of $D_N$ in powers of $g_0$. 
$D_N$ can be written as
\begin{equation}
D_N({k_1},k_2)=D_0(k_1) \,(2\pi)^4\,\delta^4({k_1}-k_2)+{\Sigma({k_1},k_2)}
\end{equation}
where $D_0(k_1)$ is the inverse propagator for zero mass fermions. 
$\Sigma({k_1},k_2)$ contains all the vertices of the overlap action;
these are made of one fermion-antifermion pair and an arbitrary number of gluons 
(up to 4 gluons for the needs of the present work). 
The much simpler case of vertices with up to 2 gluons 
(and no background) can be found in Ref.~\cite{KY}.
The expansion of $\Sigma({k_1},k_2)$ to order $g_0^4$ 
was carried out in Ref.~\cite{CP} yielding, 
after laborious analytical manipulations, all necessary
vertices for the present calculation. 
(An essential step is the expansion of $\displaystyle 1 / \sqrt{X^\dagger X}$ 
using complex analysis, which is presented in Appendix A of Ref.~\cite{CP}.)

Upon substituting the expression for $X_i$'s in the overlap vertices, 
the latter become extremely lengthy and complicated. 
For instance, the vertex with Q-Q-A-$\Psi$-$\overline{\Psi}$ consists of 9,784 terms,
while the vertex with Q-Q-A-A-$\Psi$-$\overline{\Psi}$ has 724,120 terms.


\section{RESULTS IN THE FUNDAMENTAL REPRESENTATION}

For the algebra involving lattice quantities, we make use 
of our symbolic manipulation package in Mathematica, 
with the inclusion of the additional overlap vertices. 
After a number of simplifications (involving group generators, Dirac
matrices, trigonometric identities), one must extract the
$p$-dependence of the expression corresponding to each diagram ($p$:
external momentum). Each diagram may in principle depend on $p$ as
follows:\\
\centerline{$\alpha_0 + \alpha_1\,p^2 + \alpha_2\,p^2\,\ln a^2\,p^2 +
\alpha_3\,p^2\,(\ln a^2\,p^2)^2 + \alpha_4\,(\sum_\mu p_\mu^4)/p^2 +
      {\cal{O}}(p^4, p^4\,\ln a^2\,p^2)$ }
where the coefficients $\alpha_i$ are typically 2-loop integrals with
no external momenta, which must be evaluated numerically. 
The required numerical integrations are performed by optimized Fortran
programs which are generated by our Mathematica `integrator' routine. 
Each integral is expressed as a sum over the discrete Brillouin zone
of finite lattices, with varying size $L$, and evaluated for different
values of the overlap parameter $\rho$. The average length of the
expression for each diagram, after simplifications, is about 2-3
hundred thousand terms. Finally, we extrapolate the results to
$L\to\infty$; this procedure introduces an inherent systematic error,
which we can estimate quite accurately. Infrared divergent diagrams
must be summed up before performing the extrapolation.

\smallskip
The resulting expressions for $\nu^{(1)}(p)$ and $\nu^{(2)}(p)$ (after
addition of all diagrams) are
\begin{eqnarray}
&&\nu^{(1)}(p) =  \nu^{(1)}(p)\Big|_{N_f=0} +  N_f \Biggl[k^{(1)} + 
\frac{2}{3} {\ln a^2 p^2 \over (4 \pi)^2} + {\cal O}((ap)^2) \Biggr]
\label{nu1Lattice_final}\\
\nonumber \\
&&\nu^{(2)}(p) = \nu^{(2)}(p)\Big|_{N_f=0} +  N_f \Biggl[{c^{(1,-1)}\over N} + N c^{(1,1)} +
{1\over 16 \pi^2} (3 N - {1\over N})\, {\ln a^2 p^2 \over (4 \pi)^2} + {\cal O}((ap)^2)\Biggr]
\label{nu2Lattice_final}
\end{eqnarray}
Coefficients $k^{(1)}$, $c^{(1,1)}$ and $c^{(1,-1)}$ depend on the overlap
parameter $\rho$. Their numerical results are tabulated in Tables 1
and 2 of Ref.~\cite{CP}, for different values of $\rho$ ($0<\rho<2$).

Eqs.~(\ref{nu1Lattice_final}), (\ref{nu2Lattice_final}) comply with a
number of conditions coming from comparison with continuum results and
usage of Ward Identities. Indeed:
{\bf a.} The total contribution to the gluon mass adds to zero, as expected. 
{\bf b.} The coefficients of the non-Lorentz invariant terms
($(\sum_\mu p_\mu^4)/p^2$) cancel.
{\bf c.} The terms with double logarithms correspond to the continuum counterparts. 
This has been checked diagram by diagram.
{\bf d.} Terms with single logarithms add up to their expected value, which is independent of $\rho$ 
(although the expressions per diagram are $\rho$-dependent).

In Fig. 1 we plot the 2-loop coefficients $c^{(1,-1)}$ and ($-c^{(1,1)}$) 
for different values of the overlap parameter. 
The extrapolation errors are visible for $\rho\leq 0.4$ and $\rho\geq 1.7$.
A plot of the 1-loop coefficient $k^{(1)}$ versus $\rho$ appears in
Fig. 3 of Ref.~\cite{CP}; its value lies in the interval $0.007<k^{(1)}<0.009$ for a wide
range of $\rho$ values, $0.7<\rho<1.6$, which encompasses completely
the choices made for $\rho$ in numerical simulations. 
\vskip .05cm
\centerline{\psfig{file=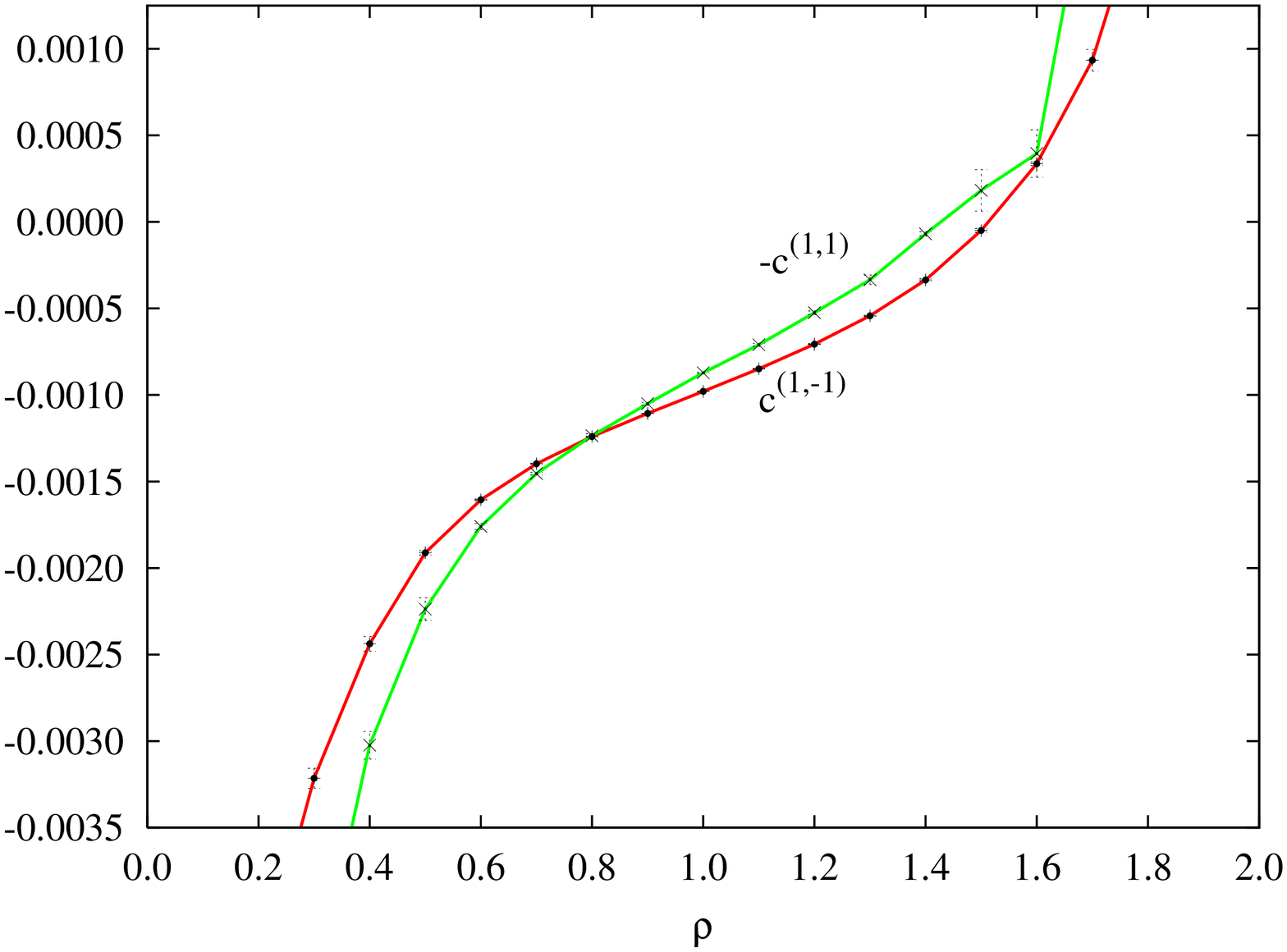,scale=.6,angle=-90}}
\vskip 2mm
\centerline{\footnotesize Fig. 1: Plot of the total 2-loop
  coefficients $c^{(1,-1)}$ and ($-c^{(1,1)}$) versus $\rho$.}
\vskip 2mm

Knowledge of the coefficients appearing in Eqs.~(\ref{nu1Lattice_final}),
(\ref{nu2Lattice_final}) allows us to derive $l_0$ (see Eq.~(\ref{l0})
below) and $l_1$ 
\begin{eqnarray}
l_1 =&& -\frac{3}{128\,N^2} +0.018127763034 - 0.007910118514\, N^2 \nonumber \\
&&+N_f \Bigg[\frac{1}{(16\pi^2)^2 N} \Big({55\over 12} -4\zeta(3) \Big) -\frac{N}{(16\pi^2)^2}{481\over 36}
-\frac{N}{8\pi^2} k^{(1)} - \Big({c^{(1,-1)}\over N} + N c^{(1,1)}  \Big)\Bigg]
\label{l1}
\end{eqnarray}
and therefore to find the 2-loop expression of $Z_g$. A direct
outcome is the final form of the 3-loop coefficient $b_2^L$ for the $\beta$-function (Eq.~(\ref{b2lrel})),
including gluonic as well as fermionic contributions; 
using Eqs. (\ref{univ_coeff_b0})-(\ref{l0}) and (\ref{l1})
\begin{eqnarray}
b_2^L =&& -{11 \over 2048\pi^2\,N} + 0.000364106020\, N -0.000092990690\, N^3 \nonumber \\
\nonumber \\
+&&N_f \Bigg[\frac{(4\pi^2-1)^2}{4 (16\pi^2)^3\,N^2} - 0.000046883436 -  000013419574\,N^2  \nonumber \\
\nonumber \\
&&\quad\quad +\frac{N_f}{(16\pi^2)^3}\Bigg( -{23\over 9\,N}+{8\zeta(3)\over 3\,N}+{37\,N\over 6}\Bigg)\nonumber \\
\nonumber \\
&&\quad\quad -\frac{(11\,N-2\,N_f)}{48\pi^2}\Big({c^{(1,-1)}\over N} + c^{(1,1)}\,N \Big) 
+ \frac{(4\,N^3+N_f-3\,N^2\,N_f)}{(16\pi^2)^2\,N}k^{(1)} \Bigg]
\label{b2_final}
\end{eqnarray}
Eq.~(\ref{b2_final}) is plotted against $\rho$ for $N=3$, $N_f=0, 2, 3$ in
Fig. 6 of Ref.~\cite{CP}. In Section V we provide the equivalent
expression for $b_2^L$ in an arbitrary representation (Eq.~(\ref{b2_L_arbitrary})).


\section{GENERALIZATION TO AN ARBITRARY REPRESENTATION}

In this Section we provide the prescription that generalizes our
results for $\nu^{(1)}$ and $\nu^{(2)}$ (Eqs. (\ref{nu1Lattice_final}),
(\ref{nu2Lattice_final})) to an arbitrary representation $r$, of dimensionality $d_r$.
For the calculation under study, only the fermion part of the action
is affected, with the link variables assuming the form
\vskip -.75cm
\begin{equation}
U_{x,\,x+\mu} = {\rm exp}(i\,g_0\,A^a_\mu(x)\,T^a_r)
\end{equation}
where $T^a_r$ denote the generators in the representation $r$, and satisfy the relations
\begin{equation}
[T^a_r,T^b_r] = i\,f^{abc}\,T^c_r,\quad \sum_aT^a_rT^a_r 
\equiv \openone\,c_r,\quad
{\rm tr}(T^a_rT^b_r)\equiv\delta^{ab}\,t_r =
\delta^{ab}\,\frac{d_r\,c_r}{N^2-1}
\label{arbirepr}
\end{equation}
In the fundamental representation $F$, one has
\begin{equation}
T^a_F\equiv T^a,\quad c_F = {N^2-1\over 2N}, \quad d_F = N,
\quad t_F={1\over2}
\label{fundrepr}
\end{equation}
Studying the color structures for each diagram with a fermionic loop
reveals the appropriate substitutions one should make, in order to
recast the results in an arbitrary representation. The generalization
prescription can be summarized in what follows. 

For the 1-loop contribution in the fundamental
representation (Eq.~(\ref{nu1Lattice_final})), the color structure is
\begin{equation}
{\rm tr}(T^aT^b) = \delta^{ab}\,t_F = {1\over2}\,\delta^{ab}
\end{equation}
Since diagrams with a closed fermion loop are always accompanied by a
factor of $N_f$, the straightforward substitution
\begin{equation}
N_f \longrightarrow N_f \cdot (2\,t_r)
\label{1loopsubstit}
\end{equation}
gives the desired results in an arbitrary representation. 

For the 2-loop fermion contribution to $\nu^{(2)}$
(Eq.~(\ref{nu2Lattice_final})), things get a bit more complicated,
because there are different types of color structures. Fortunately,
they all obey a general pattern, which will be given below.
As an example, let us consider the following diagram, and describe step by
step the extraction of the color dependence.
\vskip .75cm
\centerline{\psfig{figure=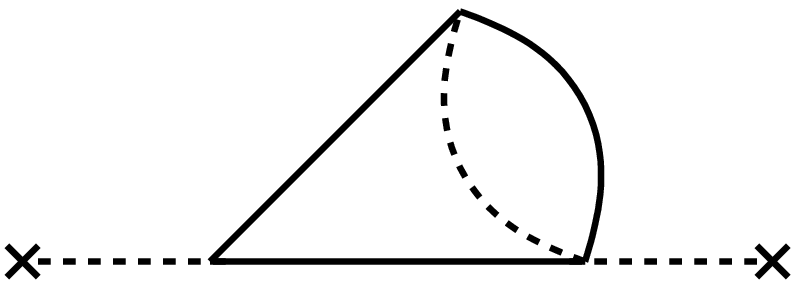,height=2truecm}}
\vskip .35cm
\noindent
\begin{minipage}[h]{4cm}
{\footnotesize{\phantom{---}}}
\end{minipage}
\hfill
\begin{minipage}[h]{11cm}
{\footnotesize {Fig. 2: A particular example of a 2-loop
    fermionic diagram. Dashed lines represent gluonic fields;
those ending on a cross stand for background gluons.
Solid lines represent fermions.}}
\end{minipage}
\hfill
\begin{minipage}[h]{4cm}
{\footnotesize{\phantom{---}}}
\end{minipage}
\vskip .75cm
\noindent
In all our diagrams, vertices with a fermion-antifermion pair and with 2 or more gluons
contain also contributions involving integrations over internal
momenta, and they are best depicted diagrammatically as non-pointlike
vertices, as shown in Fig. 3.
\vskip .75cm
\centerline{\psfig{figure=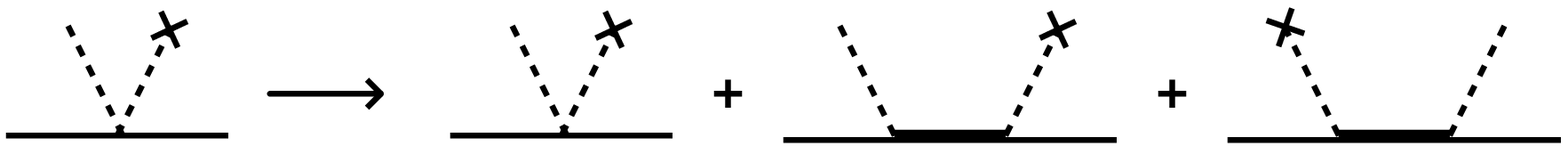,height=1.45truecm}}
\vskip .35cm
\centerline{\footnotesize {Fig. 3: The non-pointlike nature of an overlap vertex.}}
\vskip .75cm
\noindent
There is no propagator (and thus no poles) associated to the bold
fermion lines. It is important to note that the color structures
corresponding to such vertices are identical to those in ultra-local
theories (with bold lines replaced by ordinary propagators). 
The diagram of Fig. 2 actually contains two subdiagrams (Fig. 4), arising from
the non-pointlike contributions of the vertex of Fig. 3.
\vskip .75cm
\centerline{\psfig{figure=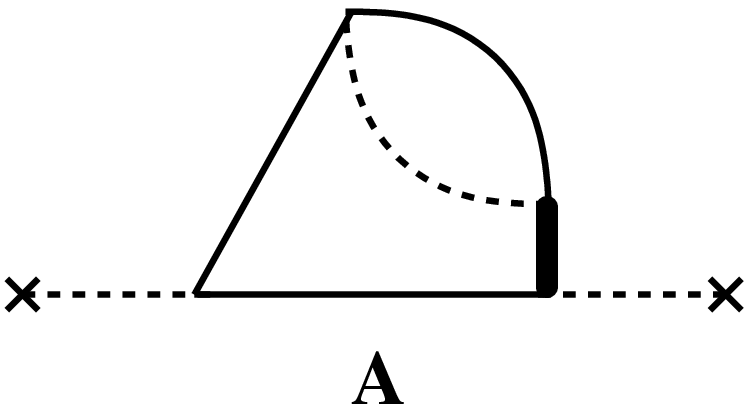,height=2.35truecm} $\quad$
  \psfig{figure=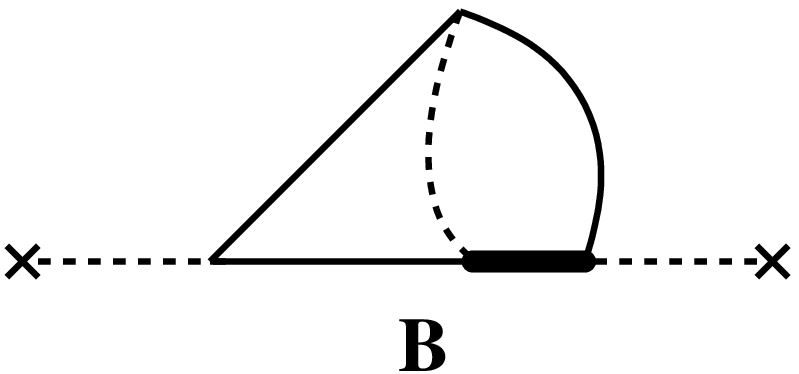,height=2.2truecm}}
\vskip .35cm
\noindent
\centerline{\footnotesize {Fig. 4: The two parts of the diagram appearing in Fig. 2.}}
\vskip .75cm
\noindent
Subdiagram A has a color dependence of the type 
\begin{equation}
{\rm tr}(T^aT^cT^cT^b)=c_r\,t_r\,\delta^{ab}
\end{equation}
($a,\,b$ color indices of the external lines), while subdiagram B has 
\begin{equation}
{\rm tr}(T^aT^cT^bT^c)=t_r\,\delta^{ab}(c_r-\frac{N}{2})
\end{equation}
Therefore, the color structure of the diagram in Fig. 2 has the form
\begin{equation}
\delta^{ab}\Bigl(\alpha\, c_r\,t_r+\beta\,t_r\,(c_r-\frac{N}{2})\Bigr)
\end{equation}
In the fundamental representation, this expression becomes
\begin{equation}
\delta^{ab}\Bigl(\alpha\, \frac{N^2-1}{4N}+\beta\,(-\frac{1}{4N})\Bigr)
\end{equation}
Thus, starting from our result for this diagram, which has the following
color dependence: $(\alpha'N + \beta'/ N)\,\delta^{ab}$, the
prescription for converting it to another representation is
\begin{eqnarray}
\left(\alpha'N + \beta'/ N\right)\,\delta^{ab} =&&
\left(4\alpha'\frac{N^2-1}{4N}-4(\alpha'+\beta')(-\frac{1}{4N})\right)\,\delta^{ab}\nonumber\\
\rightarrow &&\left(4\alpha'\,c_r\,t_r -4(\alpha'+\beta')(c_r-\frac{N}{2})\,t_r\right)\,\delta^{ab}
\label{2loopsubstit}
\end{eqnarray}
One may check that all diagrams follow the formula above.

For the computation of $b_2^L$, we need also the expressions for  $b_0,\,b_1,\,b_2$ in
an arbitrary representation
\begin{eqnarray}
b_0 &&= {1\over (4\pi)^2} 
\left({11\over 3}N-{4\over 3}t_r\,N_f\right)
\label{univ_coeff_b0} \\
b_1 &&= {1\over (4\pi)^4} \left[{34\over 3}N^2 - t_r\,N_f
  \left({20\over 3}N +4c_r \right)\right]
\label{univ_coeff_b1}\\
b_2 &&= {1\over (4\pi)^6} \left[{2857\over
    54}N^3\,+\,2\,t_r\,N_f\,\left(c_r^2-{205\over
    18}c_r\,N\,-{1415\over 54}N^2\right)
  +4\,t_r^2\,N_f^2\,\left({11\over9}c_r\,+\,{79\over54}N\right)\right]
\label{b2MS}
\end{eqnarray}
In order to calculate the ratio
$\Lambda_L/\Lambda_{\,\overline{\rm MS}}$, the quantity $l_0$ is
necessary. For overlap fermions, it equals
\begin{equation}
l_0 = {1\over 8 N} - 0.16995599 N
+ 2\,t_r\,N_f\left[ - {5\over 72 \pi^2} - k^{(1)}\right]
\label{l0}
\end{equation}
Moreover, according to the prescriptions given in Eqs.~(\ref{1loopsubstit}), (\ref{2loopsubstit}), 
the results for $\nu^{(1)}$ and $\nu^{(2)}$ become
\begin{eqnarray}
\hskip -.5cm
&&\nu^{(1)}(p) =  \nu^{(1)}(p)\Big|_{N_f=0} +  2\,t_r\,N_f\,\Biggl[k^{(1)} + 
\frac{2}{3} {\ln a^2 p^2 \over (4 \pi)^2} + {\cal O}((ap)^2) \Biggr]
\label{nu1Lattice_arbit}\\
\hskip -.5cm
\nonumber \\
\hskip -.5cm
&&\nu^{(2)}(p) = \nu^{(2)}(p)\Big|_{N_f=0} + 4\,t_r\,N_f\,\Biggl[c^{(1,1)}\frac{N}{2}
-c^{(1,-1)}\left(c_r-\frac{N}{2}\right)+\left(c_r+N\right){\ln a^2 p^2 \over (4\pi)^4} + {\cal O}((ap)^2)\Biggr]
\label{nu2Lattice_arbit}
\end{eqnarray}

Finally, Eqs.~(\ref{nu1Lattice_arbit}), (\ref{nu2Lattice_arbit}) lead to the 3-loop
coefficient of the bare $\beta$-function, which in an arbitrary
representation has the form
\newpage
\begin{eqnarray}
b_{2}^L =&& -\frac{11}{128\,(4\pi)^2}\frac{1}{N} + 0.000364106020\, N -0.000092990690\, N^3 \nonumber \\
\nonumber \\
+&&t_r\,N_f \Bigg[\frac{1}{32\,(4\pi)^2}\frac{1}{N^2}+
  \frac{c_r}{2\,(4\pi)^4}\frac{1}{N}+
  \frac{2\,c_r^2}{(4\pi)^6}-0.00011964262-0.00003220865\,c_r\,N \nonumber\\
\nonumber \\
\phantom{-}&&\phantom{t_r\,N_f}-0.00001086180\,N^2 +\frac{t_r\,N_f}{3\,(4\pi)^6}\left(
\left(\frac{184}{3} - 64\,\zeta(3)\right)\,c_r + \left(\frac{130}{3} + 32\,\zeta(3)\right)\,N \right) \nonumber\\
\nonumber \\
\phantom{-}&&\phantom{t_r\,N_f}+\frac{k^{(1)}}{32\,\pi^4}\left(-\left(c_r+N\right)\,t_r\,N_f+N^2
\right)
+\frac{c^{(1,-1)}}{24\,\pi^2}\left(2\,c_r-N\right)\,\left(-4\,t_r\,N_f+11\,N\right)\nonumber \\
\nonumber \\
\phantom{-}&&\phantom{t_r\,N_f}+\frac{c^{(1,1)}}{24\,\pi^2}N\,\left(4\,t_r\,N_f-11\,N\right)\Bigg]
\label{b2_L_arbitrary}
\end{eqnarray}


\section{ADJOINT REPRESENTATION}

As a particular application, let us focus on the adjoint
representation, $A$. The latter is encountered, e.g., in the
standard supersymmetric extension of gauge theories in terms of vector
superfields, where the gluinos are Majorana fermions in the
adjoint representation, thus similar in many respects to $N_f=1/2$
species of Dirac fermions.

The generators now take the form 
\begin{equation}
(T^a_A)_{bc} \equiv i\,f_{bac}
\end{equation} 
and the dimensionality of $A$ is $d_A=N^2-1$. Moreover,
\begin{equation}
c_A\,= N, \qquad t_A\,=\frac{d_A\,c_A}{N^2-1} = N
\label{c_A}
\end{equation} 
Eqs.~(\ref{nu1Lattice_arbit}) and (\ref{nu2Lattice_arbit}) for $\nu^{(1)}$ and $\nu^{(2)}$ now read
\begin{eqnarray}
&&\nu^{(1)}_{adj}(p) =  \nu^{(1)}(p)\Big|_{N_f=0} +  N_f\,N \Biggl[2\,k^{(1)} + 
\frac{4}{3} {\ln a^2 p^2 \over (4 \pi)^2} + {\cal O}((ap)^2) \Biggr]
\label{nu1Lattice_adj}\\
\nonumber \\
&&\nu^{(2)}_{adj}(p) = \nu^{(2)}(p)\Big|_{N_f=0} +  N_f \,N^2\Biggl[2(c^{(1,1)}-c^{(1,-1)}) +{\ln a^2 p^2 \over 32 \pi^4} + {\cal O}((ap)^2)\Biggr]
\label{nu2Lattice_adj}
\end{eqnarray}

It is interesting to find the numerical values of
$\Lambda_L/\Lambda_{\,\overline{\rm MS}}$ in the adjoint
representation (its numerical values in the fundamental representation
appear in Ref.~\cite{APV}), which can be done using our 1-loop
results for $k^{(1)}$. This ratio is defined through Eq.~(\ref{def_L}), 
where $l_0$ in this case is
\begin{equation}
l_0^{adj} = {1\over 8 N} - 0.16995599 N + 2\,N\,N_f\left[ - {5\over 72 \pi^2} - k^{(1)}\right]
\label{l0_adj}
\end{equation}
Our results for $\Lambda_L/\Lambda_{\,\overline{\rm MS}}$ are plotted in
Fig. 5 for $N=3$ and  $N_f=0,\,1/2,\,1$. One may compare Fig. 5 with
an analogous figure pertaining to fermions in the fundamental
representation (Ref.~\cite{APV}); in that case, one obtains 
$0.02\leq\Lambda_L/\Lambda_{\,\overline{\rm MS}}\leq0.025$, for $N_f=1$.
\vskip .25cm
\centerline{\psfig{figure=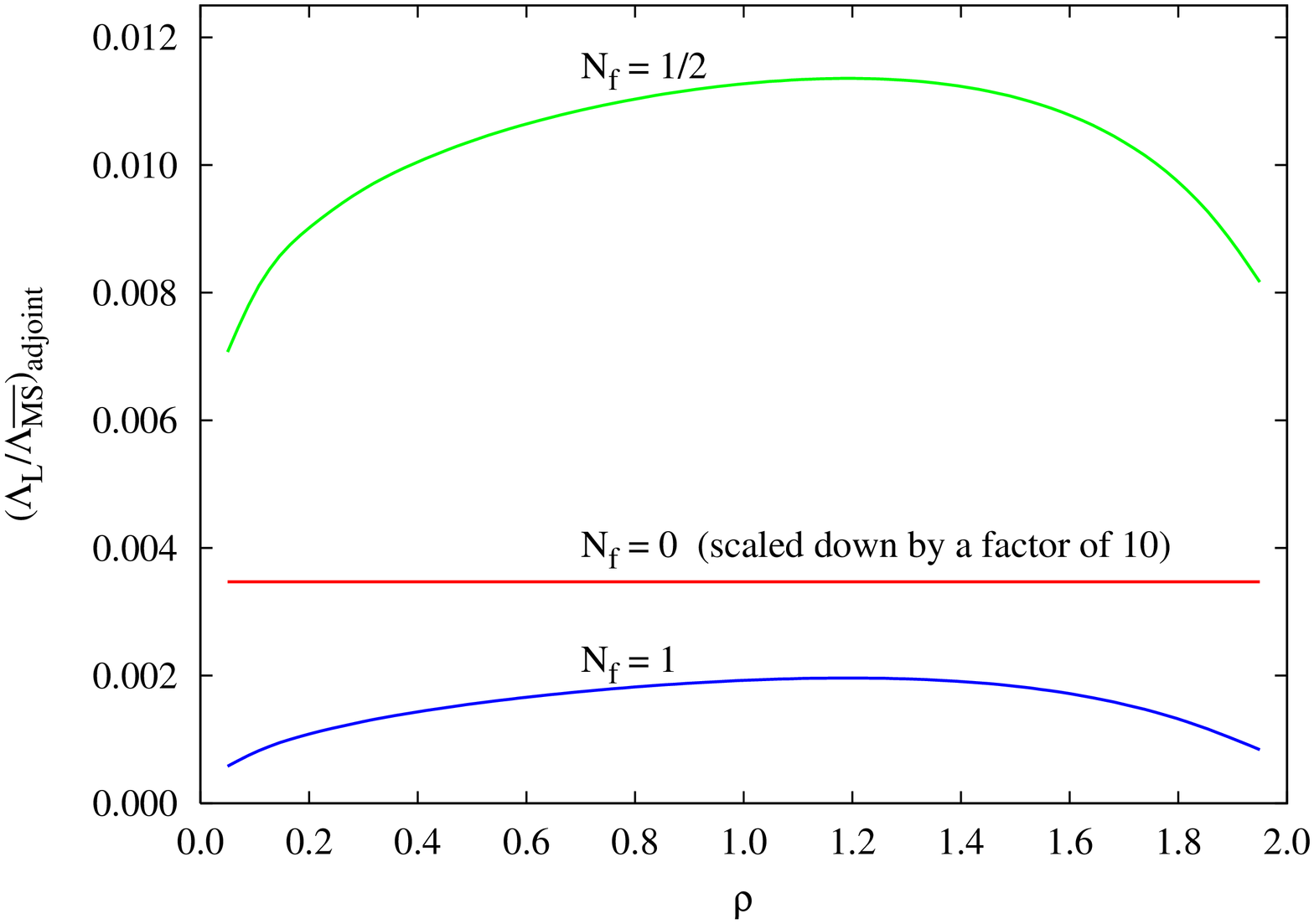,scale=.6,angle=-90}}

\vskip .25cm
\noindent
{\footnotesize {Fig. 5: The  $\rho$ dependence of the ratio
    $\Lambda_L/\Lambda_{\,\overline{\rm MS}}$ in the adjoint
    representation is plotted for $N=3$ and
    $N_f=0$ (horizontal line {\it{scaled down by a factor of}} 10
    {\it{from its value}} 0.034711), $N_f=1/2$ and $N_f=1$.}}
\vskip .25cm

\bigskip\noindent
{\bf Acknowledgements:} This work is supported in part by the
Research Promotion Foundation of Cyprus (Proposal Nr: $\rm ENI\Sigma X$/0505/45).

\newpage

\end{document}